\documentclass[twocolumn,showpacs,prl,amsmath,amssymb]{revtex4}
\usepackage{graphicx}% Include figure files
\usepackage{dcolumn}% Align table columns on decimal point
\usepackage{bm}% bold math

\newcommand{\ket}[1]{|\hspace{0.5pt}#1\hspace{0.5pt}\rangle}

\newcommand{\bra}[1]{\langle\hspace{0.5pt}#1\hspace{0.5pt}|}

\newcommand{\op}[1]{\operatorname{#1}}

\def \qed {\hfill \rule{2mm}{2mm}\vspace{3mm}}

\newtheorem{theorem}{Theorem}
\newtheorem{lemma}[theorem]{Lemma}

\begin{document}

%\preprint{}

\title{Many copies may be required for entanglement distillation}

\author{John Watrous}
\email{jwatrous@cpsc.ucalgary.ca}
\affiliation{
  Department of Computer Science\\
  University of Calgary\\
  Calgary, Alberta, Canada~T2N 1N4}

\date{April 29, 2004}

\begin{abstract}
A mixed quantum state $\rho$ shared between two parties is said to be
{\em distillable} if, by means of a protocol involving only local quantum
operations and classical communication, the two parties can transform some
number of copies of $\rho$ into a single shared pair of qubits
having high fidelity with the maximally entangled state
$\ket{\phi^+} = (\ket{00} + \ket{11})/\sqrt{2}$.
In this paper it is proved that there exist states that are distillable,
but for which an arbitrarily large number of copies is required before
any distillation procedure can produce a shared pair of qubits with even
a small amount of entanglement.
Specifically, for every positive integer $n$ there exists
a state $\rho$ that is distillable, but given $n$ or fewer copies
of $\rho$ every distillation procedure outputting a single shared pair
of qubits will output those qubits in a separable (i.e., unentangled) state.
Essentially all previous examples of states proved to be distillable were such
that some distillation procedure could output an entangled pair of qubits
given a single copy of the state in question.
\end{abstract}


%\keywords{}

\maketitle

%=============================================================================%

\section{Introduction}

Entanglement represents an important resource in quantum information theory.
For example, by means of quantum teleportation \cite{BennettB+93}, entanglement
shared between two parties that may only send classical information to one
another allows the parties to exchange quantum information.
Superdense coding \cite{BennettW92}, which allows one qubit of quantum
communication to transmit two classical bits of communication using prior
entanglement, is another example where entanglement is used as a resource.
{}From the point of view of such protocols, a shared pair of qubits in the
state $\ket{\phi^+} = (\ket{00} + \ket{11})/\sqrt{2}$ (or any other maximally
entangled state) represents one unit of entanglement, known as an {\em e-bit}.
For instance, at the cost of one e-bit plus two classical bits of
communication, quantum teleportation allows for the transmission of one qubit
of information.

Suppose that two parties, Alice and Bob, would like to perform quantum
teleportation or some other protocol based on entanglement, but instead
of sharing copies of the state $\ket{\phi^+}$ they share copies of
some other quantum state $\rho$.
For instance, $\rho$ may represent a noisy copy of $\ket{\phi^+}$ that
will not allow for sufficiently accurate transmission of quantum information
by Alice and Bob's standards, or $\rho$ may be a strange quantum state
that is entangled but has no resemblance whatsoever to $\ket{\phi^+}$.
The process of {\em entanglement distillation}, first considered by
Bennett, et.al.~\cite{BennettBernstein+96}, addresses this situation---by means
of some protocol allowing Alice and Bob to perform only local quantum
operations and to communicate classically (an {\em LOCC protocol}, for short),
some number of copies of $\rho$ may be transformed into some (possibly
smaller) number of copies of $\ket{\phi^+}$ with high accuracy.
When it is possible for Alice and Bob to transform one or more copies
of $\rho$ into at least one copy of $\ket{\phi^+}$ with high accuracy
in this way, $\rho$ is said to be {\em distillable}.

Some states $\rho$ are distillable and some are not.
In the case where $\rho$ is a pure, entangled state, distillation is
always possible \cite{BennettBernstein+96}; even if $\rho$ has a very small
amount of entanglement, sufficiently many copies of $\rho$ will allow copies of
$\ket{\phi^+}$ to be distilled with high accuracy.
Similarly, if $\rho$ is a mixed state of exactly two qubits,
$\rho$ being distillable is equivalent to $\rho$ being entangled
\cite{BennettBrassard+96,HorodeckiH+97}.
In the general case for mixed states, however, there are examples of states
that are entangled but are not distillable \cite{HorodeckiH+98}.
Such state are known as {\em bound-entangled states}.

All currently known examples of bound entangled states have the property
that the partial transpose of the density operator of the state in question
is positive semidefinite.
States of this sort are called PPT (positive partial transpose) states for
short.
While every PPT state is undistillable, the converse is not known to
hold, and it is a central open question in the theory of entanglement
to determine whether or not this is the case.
More generally speaking, there is no effective procedure known to determine
whether a given state is distillable or not.
For a certain range of parameters, Werner states have been conjectured to be
examples of bound entangled non-PPT states \cite{DiVincenzoS+00, DurC+00}.

Some of the difficulty in understanding entanglement distillation may be
attributed to the fact that, by definition, an arbitrary number of copies of
the state in question may be used in the distillation process.
Suppose that instead of having an unlimited number of copies of a given
bipartite state $\rho$, Alice and Bob have some fixed number
of copies that they wish to subject to distillation.
One says that $\rho$ is {\em $n$-distillable} if there exists an LOCC protocol
whereby Alice and Bob can convert $n$ copies of $\rho$ to a shared pair of
qubits that is entangled.
It should be stressed that this definition places no restriction on
the amount of entanglement of the shared pair of qubits output by the
procedure; it only requires that this pair of qubits are in some
entangled (i.e., non-separable) state.
Note, however, that a necessary and sufficient condition for a state $\rho$ to
be distillable is that $\rho$ is $n$-distillable for some $n$.
This is because a large number of copies of $\rho$ can be collected into
groups of size $n$, the distillation procedure used to produce
entangled pairs of qubits, then these pairs of entangled qubits further
distilled using the procedure of Ref.~\cite{HorodeckiH+97}.
For pure states and for mixed states on a single shared pair of qubits,
distillability and 1-distillability are equivalent.

The main result of this paper establishes that for any given value of $n$,
there exist states that are distillable but not $n$-distillable.
This was not previously observed even for the case $n=1$.
The dimension of such states does not need to depend on $n$;
9 $\otimes$ 9 dimensions are sufficient for the existence of such states for
all values of $n$.

\begin{theorem}
\label{theorem:main}
For any choice of integers $d\geq 3$ and $n\geq 1$, there exists
a $d^2\otimes d^2$ bipartite mixed quantum state that is distillable but not
$n$-distillable.
\end{theorem}

\noindent
{\bf Remark.}
It should be noted that for the particularly simple case of $n=1$,
this theorem follows from results proved in Ref.~\cite{ShorS+01}.
Specifically, it is implicit in that paper that there exist states $\rho$
and $\xi$ that are not 1-distillable (and in fact $\xi$ is not distillable
at all), but such that $\rho\otimes\xi$ is 1-distillable.
Assuming without loss of generality that these are states of systems of
equal size, it follows that the state
$\frac{1}{2}\ket{00}\bra{00}\otimes \rho + 
\frac{1}{2}\ket{11}\bra{11}\otimes \xi$ is 2-distillable but not 1-distillable.
A similar example can be derived from the results of Ref.~\cite{VollbrechtW02}.
It is not clear, however, that this construction can be extended beyond the
case $n = 1$.

%=============================================================================%

\section{Preliminaries}
\label{sec:preliminaries}

Let $\mathcal{A}$ and $\mathcal{B}$ be Hilbert spaces.
A vector $\ket{\psi}\in\mathcal{A}\otimes\mathcal{B}$ is said to have
{\em Schmidt rank} $k$ if
\[
\op{rank}(\op{tr}_{\mathcal{A}}\ket{\psi}\bra{\psi}) = k,
\]
where $\op{tr}_{\mathcal{A}}: \mathrm{L}(\mathcal{A}\otimes\mathcal{B})
\rightarrow\mathrm{L}(\mathcal{B})$ denotes the partial trace.
Given a linear operator $X\in\mathrm{L}(\mathcal{A}\otimes\mathcal{B})$,
the partial transpose over $\mathcal{A}$ applied to $X$ is denoted
$T_{\mathcal{A}}(X)$.
Transposition must be taken with respect to a particular basis of
$\mathcal{A}$, which is always assumed to be a given standard basis in this
paper.

The following fact, first proved in Ref.~\cite{HorodeckiH+98}, allows
entanglement distillation to be characterized without reference to LOCC
transformations.
A density matrix $\rho$ acting on $\mathcal{A}\otimes\mathcal{B}$ is
$1$-distillable if and only if there exists some Schmidt rank 2 vector
$\ket{\psi}\in\mathcal{A}\otimes\mathcal{B}$ for which
\[
\bra{\psi} T_{\mathcal{A}}(\rho)\ket{\psi} < 0,
\]
and $\rho$ is $n$-distillable if $\rho^{\otimes n}$ is 1-distillable.
If $\rho$ is $n$-distillable for some integer $n\geq 1$, then
$\rho$ is distillable, otherwise $\rho$ is undistillable.
It is convenient that this characterization holds regardless of whether
the state $\rho$ is normalized.
Consequently, normalization factors for density matrices will often be
ignored in this paper.

A convention that will be followed throughout this paper is that the
Hilbert space $\mathcal{A}$ always refers to Alice's part of a given system
and $\mathcal{B}$ refers to Bob's part.
Schmidt rank and any reference to distillation will generally be with respect
to this partition.
Different symbols, such as $\mathcal{F}$, $\mathcal{G}$, $\mathcal{H}$, etc.,
will be used to refer to Hilbert spaces of systems not necessarily shared
between Alice and Bob in this way in order to avoid confusion.

Let $\mathcal{F}$ and $\mathcal{G}$ be $d$-dimensional Hilbert spaces, and
let $\{\ket{1},\ldots,\ket{d}\}$ be the standard basis for both of these
spaces.
Four projection operators on $\mathcal{F}\otimes\mathcal{G}$ will play an
important role in this paper.
The first two projections are
\begin{eqnarray*}
P & = & \ket{\Phi}\bra{\Phi},\\
Q & = & I - \ket{\Phi}\bra{\Phi},
\end{eqnarray*}
where
\[
\ket{\Phi} = \frac{1}{\sqrt{d}}\sum_{i=1}^d\ket{i}\ket{i}.
\]
The other two projections are
\begin{eqnarray*}
R & = & \frac{1}{2}(I - F),\\[2mm]
S & = & \frac{1}{2}(I + F)\:=\:I - R,
\end{eqnarray*}
where
\[
F = \sum_{1\leq i,j\leq d}\ket{i}\bra{j}\otimes\ket{j}\bra{i}.
\]
The projection $R$ is the projection onto the antisymmetric subspace
of $\mathcal{F}\otimes\mathcal{G}$, while $S$ is the projection onto
the symmetric subspace of $\mathcal{F}\otimes\mathcal{G}$.
The following relations among these projections and the partial transpose
hold.
\begin{eqnarray*}
T_{\mathcal{F}}(P) & = & -\frac{1}{d} R + \frac{1}{d} S,\\
T_{\mathcal{F}}(Q) & = & \frac{d+1}{d} R + \frac{d-1}{d} S,\\
T_{\mathcal{F}}(R) & = & - \frac{d-1}{2} P + \frac{1}{2} Q,\\
T_{\mathcal{F}}(S) & = & \frac{d+1}{2} P + \frac{1}{2} Q.
\end{eqnarray*}

%=============================================================================%

\section{Proof of Theorem~\ref{theorem:main}}
\label{sec:proof}

Consider a system with four $d$-dimensional components, two in Alice's
possession and two in Bob's possession.
It will be convenient to refer to these systems as {\em quantum registers}
$X_1, \ldots, X_4$ with corresponding Hilbert spaces
$\mathcal{H}_1,\ldots,\mathcal{H}_4$.
The standard basis for these spaces will be taken to be
$\{\ket{1},\ldots,\ket{d}\}$.
Later it will be necessary to consider systems with more registers, which
will be labeled similarly and will have corresponding Hilbert spaces labeled
similarly.
In all cases, it is assumed that Alice possesses the odd-numbered registers
and Bob possesses the even-numbered registers.
When necessary, the tensor product structure of various operators will be
indicated by subscripts that index these systems.
For example, the projection $R$ on $\mathcal{H}_1 \otimes \mathcal{H}_2$
tensored with the projection $S$ on $\mathcal{H}_3 \otimes \mathcal{H}_4$
is denoted $R_{1,2}\otimes S_{3,4}$.

Define the (unnormalized) state $\rho(\varepsilon)$ as follows:
\[
\rho(\varepsilon) = \frac{d+1+\varepsilon}{d-1}R_{1,2}\otimes R_{3,4}
+ S_{1,2}\otimes S_{3,4}.
\]
Theorem~\ref{theorem:main} will follow from these two lemmas:

\begin{lemma}
\label{lemma:n-copy}
For any integers $d\geq 3$ and $n\geq 1$, there exists
a real number $\varepsilon > 0$ such that $\rho(\varepsilon)$ is not
$n$-distillable.
\end{lemma}

\begin{lemma}
\label{lemma:distill}
For every $d\geq 3$ and $\varepsilon > 0$, the state $\rho(\varepsilon)$
is distillable.
\end{lemma}

\noindent
{\bf Proof of Lemma~\ref{lemma:n-copy}.}
Let $\mathcal{A} = \mathcal{H}_1\otimes\mathcal{H}_3$ and
$\mathcal{B} = \mathcal{H}_2\otimes\mathcal{H}_4$.
The partial transpose of $\rho(\varepsilon)$ is
\begin{equation}
\label{eq:ptrho}
T_{\mathcal{A}}(\rho(\varepsilon)) =
\frac{1}{4}\left(
\mu P\otimes P 
-\varepsilon\, P\otimes Q 
-\varepsilon\, Q\otimes P 
+ \lambda\, Q\otimes Q 
\right)
\end{equation}
where
$\mu = (d+1)^2 + (d+1+\varepsilon)(d-1)$ and
$\lambda = 1 + \frac{d+1+\varepsilon}{d-1}$.
The partial transpose of $n$ copies of $\rho(\varepsilon)$
can be expressed as
\[
\left(T_{\mathcal{A}}\left(\rho(\varepsilon)\right)\right)^{\otimes n}
=
\frac{1}{4^n}\sum_{x\in\{0,1\}^{2n}} \alpha(x) \Pi_x,
\]
where $\Pi_0 = P$, $\Pi_1 = Q$,
$\Pi_x = \Pi_{x_1}\otimes\cdots\otimes\Pi_{x_{2n}}$ for $x\in\{0,1\}^{2n}$,
and each coefficient $\alpha(x)$ is easily determined by the
equation (\ref{eq:ptrho}) above.
In particular, these coefficients satisfy
$\alpha(1^{2n}) = \lambda^n$
$\alpha(0^{2n}) = \mu^n$, 
and $|\alpha(x)| \leq \varepsilon\mu^{n-1}$ for all
$x\not\in\{0^{2n},1^{2n}\}$.

Suppose that
$\ket{\psi}\in\mathcal{A}^{\otimes n} \otimes \mathcal{B}^{\otimes n}$
is a unit vector having Schmidt rank equal to 2.
Then
\[
\bra{\psi} Q^{\otimes 2n}\ket{\psi} \geq \left(1-\frac{2}{d}\right)^{2n}.
\]
This inequality is proved in D\"ur, et.al.~\cite{DurC+00} for the case
$d = 3$, and generalizes to arbitrary $d$ without complications.
It follows that
\[
\bra{\psi}T_{\mathcal{A}}\left(\rho(\varepsilon)\right)^{\otimes n}\ket{\psi}
\geq \frac{\lambda^n}{4^n}\left(1 - \frac{2}{d}\right)^{2n}
- \varepsilon\mu^{n-1}.
\]
Because $\lambda$ and $\mu$ can be lower bounded and upper bounded,
respectively, by positive reals not depending on $\varepsilon$, it follows
that the above quantity is positive for sufficiently small $\varepsilon >0$.
For such a choice of $\varepsilon$, it is therefore the case that
$\rho(\varepsilon)$ is not $n$-distillable.
\qed

\noindent
{\bf Proof of Lemma~\ref{lemma:distill}.}
It is assumed that Alice and Bob have an unbounded supply of copies of
$\rho(\varepsilon)$.
Alice and Bob will iterate a particular process involving eight $d$-dimensional
registers $X_1,\ldots,X_8$ with corresponding Hilbert spaces
$\mathcal{H}_1,\ldots,\mathcal{H}_8$.
As before, it is assumed that Alice possesses the odd-numbered registers
and Bob possesses the even-numbered registers.

Suppose at some instant that the registers $X_1,\ldots,X_4$ contain the
state
\[
\alpha R_{1,2}\otimes R_{3,4} + S_{1,2}\otimes S_{3,4}
\]
for some $\alpha\geq 0$, while registers $X_5,\ldots,X_8$ contain
a copy of $\rho(\varepsilon)$, i.e., 
\[
\frac{d+1+\varepsilon}{d-1}R_{5,6}\otimes R_{7,8}
+ S_{5,6}\otimes S_{7,8}.
\]
Alice measures the pair $(\mathrm{X}_1,\mathrm{X}_5)$ with respect to the
measurement described by $\{P,Q\}$ and Bob does likewise with the pair
$(\mathrm{X}_2,\mathrm{X}_6)$.
The process being iterated fails if either of the measurement outcomes
does not correspond to the projection $P$.
In case they both obtain an outcome corresponding to projection $P$,
they discard the registers on which they performed the measurements, which
leaves the 4 registers $(X_3,X_4,X_7,X_8)$ in the state
\begin{align*}
& \alpha\frac{d+1+\varepsilon}{d-1}
\op{tr}\left((P_{1,5}\otimes P_{2,6})(R_{1,2}\otimes R_{5,6})\right)
R_{3,4}\otimes R_{7,8}\\
& + \alpha \op{tr}\left((P_{1,5}\otimes P_{2,6})(R_{1,2}\otimes S_{5,6})\right)
R_{3,4}\otimes S_{7,8} \\
& + \frac{d+1+\varepsilon}{d-1}
\op{tr}\left((P_{1,5}\otimes P_{2,6})(S_{1,2}\otimes R_{5,6})\right)
S_{3,4}\otimes R_{7,8}\\
& + \op{tr}\left((P_{1,5}\otimes P_{2,6})(S_{1,2}\otimes S_{5,6})\right)
S_{3,4}\otimes S_{7,8}.
\end{align*}
One may calculate that
%\begin{eqnarray*}
%\op{tr}\left((P_{1,3}\otimes P_{2,4})(I_{1,2}\otimes I_{3,4})\right) & = &
%\op{tr}\left((P_{1,3}\otimes P_{2,4})(F_{1,2}\otimes F_{3,4})\right)\:=\: 1,\\
%\op{tr}\left((P_{1,3}\otimes P_{2,4})(I_{1,2}\otimes F_{3,4})\right) & = &
%\op{tr}\left((P_{1,3}\otimes P_{2,4})(F_{1,2}\otimes I_{3,4})\right) \:=\:
%\frac{1}{d},
%\end{eqnarray*}
%and therefore
\begin{align*}
\op{tr}\left((P_{1,5}\otimes P_{2,6})(R_{1,2}\otimes R_{5,6})\right) = &
\frac{d-1}{2d},\\[1mm]
\op{tr}\left((P_{1,5}\otimes P_{2,6})(R_{1,2}\otimes S_{5,6})\right) = &
0,\\[1mm]
\op{tr}\left((P_{1,5}\otimes P_{2,6})(S_{1,2}\otimes R_{5,6})\right) = &
0,\\[1mm]
\op{tr}\left((P_{1,5}\otimes P_{2,6})(S_{1,2}\otimes S_{5,6})\right) = &
\frac{d+1}{2d},
\end{align*}
and therefore the state of the registers $(X_3,X_4,X_7,X_8)$ above is
\[
\frac{d+1}{2d} \left(\alpha\left(1 + \frac{\varepsilon}{d+1}\right)
R_{3,4}\otimes R_{7,8} + S_{3,4}\otimes S_{7,8}\right).
\]

Now, based on this process, Alice and Bob will distill their copies
of $\rho(\varepsilon)$ as follows.
They begin with $(X_1,\dots,X_4)$ and $(X_5,\dots,X_8)$ each containing a
copy of $\rho(\varepsilon)$, and the above iteration is performed.
If it is successful, they relabel registers $(X_3,X_4,X_7,X_8)$
as $(X_1,X_2,X_3,X_4)$ and initialize $(X_5,\dots,X_8)$ with a new
copy of $\rho(\varepsilon)$.
Otherwise, if it is not successful, they start the entire process over with
both $(X_1,\dots,X_4)$ and $(X_5,\dots,X_8)$ initialized to
$\rho(\varepsilon)$.
This process is repeated until a number $k$ of consecutive successes
has been achieved that satisfies
\[
\frac{d + 1 + \varepsilon}{d-1}
\left(1 + \frac{\varepsilon}{d+1}\right)^k > 3.
\]
This eventually happens with probability 1.
At a point when it has happened, the registers $(X_1,X_2,X_3,X_4)$ will contain
a state of the form $\alpha R_{1,2}\otimes R_{3,4} + S_{1,2}\otimes S_{3,4}$
for $\alpha > 3$.

It remains to prove that for $\alpha > 3$ the state
\[
\alpha R_{1,2}\otimes R_{3,4} + S_{1,2}\otimes S_{3,4}
\]
is 1-distillable.
To see this, consider the Schmidt rank 2 vector
\[
\ket{\phi} = \ket{1}_1\ket{2}_2\left(\ket{1}_3\ket{1}_4 + 
\ket{2}_3\ket{2}_4\right).
\]
Because
\begin{align*}
T_{\mathcal{A}}(\alpha & R_{1,2}\otimes R_{3,4} + S_{1,2}\otimes S_{3,4})\\
= & \frac{\alpha + 1}{4} I_{1,2}\otimes I_{3,4}
-\frac{(\alpha - 1)d}{4} I_{1,2}\otimes P_{3,4} \\
& -\frac{(\alpha - 1)d}{4} P_{1,2}\otimes I_{3,4}
+ \frac{(\alpha+1)d^2}{4} P_{1,2}\otimes P_{3,4},
\end{align*}
it follows that
\[
\bra{\phi}T_{\mathcal{A}}(\alpha R_{1,2}\otimes R_{3,4} + 
S_{1,2}\otimes S_{3,4})\ket{\phi}
= \frac{3 - \alpha}{2} < 0.
\]
This completes the proof.
\qed

%=============================================================================%

\section{Discussion}

Theorem \ref{theorem:main} establishes a counter-intuitive property of
entanglement distillation, which is that entanglement distillation is
nonlinear with respect to the number of copies used in the distillation
process.
It is curious that there exist, for instance, examples of quantum states
$\rho$ such that $10^6$ copies of $\rho$ are not sufficient for a single
shared pair of non-separable qubits to be distilled, but with many more copies
of $\rho$ near-perfect e-bits can be distilled.

As discussed in the introduction, no effective procedure is known
that determines whether or not a given bipartite state is distillable.
This paper rules out the possibility that distillability
is equivalent to $n$-distillability for some finite value of $n$, and
therefore implies the characterization for $n$-distillability introduced
in Ref.~\cite{HorodeckiH+98} and discussed above does not extend to an
effective test for distillability in any obvious way.

Finally, the result proved here has implications to the conjecture
of Refs.~\cite{DiVincenzoS+00,DurC+00} concerning the distillability of
Werner states for certain ranges of parameters.
More specifically, the (unnormalized) family of Werner states
$\sigma_W(\alpha) = S + \alpha R$ in $d\otimes d$ dimensions are readily
seen to be non-PPT states for $(d+1)/(d-1) < \alpha$, and 1-distillable if
and only if $\alpha > 3$.
The conjecture of Refs.~\cite{DiVincenzoS+00,DurC+00} is that
$\sigma_W(\alpha)$ is undistillable for $\alpha \leq 3$,
which would imply that the PPT states are a proper subset of the
undistillable states.
One of the pieces of evidence presented in support of this conjecture
was that for every positive integer $n$, there exists some value of
$\alpha > (d+1)/(d-1)$ for which $\sigma_W(\alpha)$ is not
$n$-distillable.
(Indeed, the proof of Lemma~\ref{lemma:n-copy} above proceeds
along similar lines to the proof of this fact from Ref.~\cite{DurC+00}.)
The present paper certainly does not refute this conjecture,
but does call into question the evidence just discussed.
In particular, the states $\rho(\varepsilon)$ defined in the proof of
Theorem~\ref{theorem:main} possess essentially the same property of being
neither PPT nor $n$-distillable for some choice of $\varepsilon$, but
nevertheless are distillable.
Perhaps this fact may shed some light on the question of whether or not
non-PPT states can always be distilled.

%=============================================================================%

\subsection*{Acknowledgments}

Thanks to John Smolin, Ashish Thapliyal, and the anonymous referees for
very helpful comments and corrections.
This research was supported by Canada's NSERC and the Canada Research Chairs
program.

%=============================================================================%

%\bibliography{references}

\end{document}